\documentclass[aps,prb,onecolumn,showpacs]{revtex4}
\usepackage{amsmath}
\usepackage{graphics}
\usepackage{epsfig}
\usepackage{color}
\usepackage{amsfonts,amsmath,mathrsfs}
\usepackage{slashed}
\usepackage{hyperref}
\usepackage{epsfig}
\usepackage{latexsym}
\usepackage{bbm}
\usepackage{amssymb}
\usepackage{graphicx}
\usepackage{dcolumn}
\usepackage{bm}

\begin{document}

\title{Superconductivity in Mutual Chern-Simons Gauge Theory}
\author{Peng Ye\footnote{Electronic address: yep07@mails.tsinghua.edu.cn}, Long Zhang, and Zheng-Yu Weng}
\affiliation{Institute for Advanced Study, Tsinghua University, Beijing, 100084, People's Republic of China}


\begin{abstract}
In this work, we present a topological characterization of superconductivity
in a prototype electron fractionalization model for doped Mott insulators.
In this model, spinons and holons are coupled via the mutual Chern-Simons
gauge fields. We obtain a low-lying effective description of the collective current fluctuations by integrating out the matter fields, which replaces the conventional Ginzburg-Landau action to describe the
generalized rigidity of superconductivity. The superconducting phase coherence
is essentially characterized by a topological order parameter related to
a Gaussian linking number, and an experiment is proposed to probe this
topological property. We further show that a gauge-neutral fermionic mode can naturally emerge in this model, which behaves like a Bogoliubov quasiparticle.
\end{abstract}

\date{{\small \today}}
\maketitle


\section{Introduction}

The microscopic nature of high-$T_{c}$ superconductivity in the cuprates is
still under heavy debate after more than two decades of intensive studies.
It has been widely accepted that the cuprates may be properly considered as
doped Mott insulators and the superconductivity arises after the
antiferromagnetic (AFM) long range order is destroyed by doping. The
challenge is that a typical doped Mott insulator is a strongly correlated
electron system in which two electrons staying at the same lattice site will
cost a huge energy as compared to other energy scales like the hopping
energy. With the failure of conventional perturbative many-body methods, to
tackle a doped Mott insulator one usually resorts to either the so-called
Gutzwiller projected trial wavefunction or constructing an effective
low-energy theory \cite{Lee06}. In the latter approach, the electron
fractionalization \cite{Lee06,Kivelson87,Zou88,Senthil00,Weng07} is
often utilized in order to implement some key features of a doped Mott
insulator. Indeed, the fractionalization due to the Mottness is most
transparent at half-filling where only the spin degrees of freedom of the
electrons are left, with the charge part being totally frozen out. Only by
doping can charge carriers emerge which  account for a small portion of
the total number of the electrons at low doping.

The real challenge is that there exist many different ways of introducing
fractionalized elementary objects to implement the same local Mottness,
i.e., the no double occupancy constraint. These elementary particles are
generally further coupled to some emergent gauge field(s) due to the
uncertainty in the fractionalization \cite{Lee06}. In the lack of precise
energetic comparison, a full self-consistent development of a potential
fractionalization/gauge theory becomes essential in order for it to be
convincingly verified or falsified by experiment as well as by general
theoretical considerations. In principle, an inappropriate fractionalization
would always lead to strong gauge fluctuations rendering the effective
theory intractable. Therefore, a stable, well-controlled gauge theory
description will be a highly desired candidate for the correct low-energy
theory of cuprate superconductors.

We will consider a particular electron fractionalization description with a
mutual Chern-Simons gauge structure \cite{Kou05,Ye1,Ye2}. Specifically, an
electron in this theory is \textquotedblleft
fractionalized\textquotedblright \ into an $S=1/2$ bosonic spinon and a
charge $+e$ bosonic holon, which are minimally coupled to U(1) gauge fields $%
A^{h}$ and $A^{s}$, which do not have their own dynamics, but are
\textquotedblleft entangled\textquotedblright \ by a mutual Chern-Simons
topological term
\begin{equation}
\mathcal{L}_{\text{MCS}}=\frac{i}{\pi }\epsilon ^{\mu \nu \lambda }A_{\mu
}^{s}\partial _{\nu }A_{\lambda }^{h}\text{.}  \label{LMCS}
\end{equation}%
which is (2+1)D realization of the topological BF theory.\cite{Birmingham91} Physically it simply means that the spinons and holons will perceive each
other as quantized $\pi $ fluxoids \cite{Weng97}. In the limit of zero hole
concentration, the spinon condensation can naturally lead to an AFM
long-range ordered state. At finite doping, the mutual Chern-Simons term in
Eq. (\ref{LMCS}) captures \cite{Kou05,Ye1} a precise \textquotedblleft
statistical sign structure\textquotedblright \ identified \cite{Weng97,Wu08}
in the $t$-$J$ model. By the $\pi $ fluxoids carried by the mobile holons,
the AFM state will be destroyed and eventually a superconducting will set in
with the holon condensation, and a zero-temperature phase diagram as a
function of doping concentration has been recently discussed in Refs. \onlinecite%
{Ye1,Ye2}. Finally, there are other quantum systems \cite%
{DST96,Galitski05,Cho11,Xu09} which have the mutual Chern-Simons gauge
structures in different contexts.

In this paper, we focus on the superconducting state to develop an effective description based on the mutual Chern-Simons gauge theory in Sec.
II. Instead of a conventional Ginzburg-Landau theory in description of the
fluctuations of the pairing order parameter, we obtain a gauge description
of the low-lying collective fluctuations of charge and spin currents, which
are essentially the phase fluctuations of the superconducting order
parameter. We show that the Meissner effect in such a superconductor is
basically controlled by a spin gap. In Sec. III, we further\ identify a
non-BCS-type order parameter $\mathcal{I}$ that characterizes the SC phase
coherence. Here $\mathcal{I}$ is related to the correlation function of two
Wilson loops (of $A^{s}$ and $A^{h}$) and is topological as it merely
depends on the Gaussian linking number under a continuous deformation of the
loops. Such a relation is similar to the connection between the Chern-Simons
theory and Knot theory first studied in Ref. \onlinecite{Witten89}. But by coupling
to the matter fields, we show that the spin correlations strongly
renormalize $\mathcal{I}$ such that the phase coherence terminates at a
critical point where the spin gap vanishes. In Sec. III C, we shall propose
a direct experimental probe of such a topological nature of the gauge
interaction between the spin and charge degrees of freedom. Finally, in Sec.
III D, we show that a fermionic excitation can emerge in the superconducting
phase, which is gauge neutral and thus coherent, resembling a Bogoliubov
quasiparticle.

\section{Effective Field Theory Description of Superconductivity}

\subsection{Lattice mutual Chern-Simons theory}

The lattice field theory formulation of the phase string theory\cite{Weng07}
for a doped Mott insulator has been recently constructed in Refs. \onlinecite{Ye1,Ye2}
in terms of partition function:
\begin{equation}
\mathcal{Z}=\sum_{\{ \mathscr{N}^{s},\mathscr{N}^{h}\}}\int
D[A^{s},A^{h}]D[h^{\dagger },h,b^{\dagger },b]e^{-S}  \label{partition}
\end{equation}%
in which the action $S=\sum_{x}\mathcal{L}$ ($x$ denotes spacetime
coordinates) with
\begin{equation}
\mathcal{L}=\mathcal{L}_{h}+\mathcal{L}_{s}+\mathcal{L}_{\mathrm{MCS}}\text{
.}  \label{S}
\end{equation}

Here the first two terms, $\mathcal{L}_{h}$ and $\mathcal{L}_{s}$, are the
Lagrangian densities for the matter fields, holon $h_{i}$ and spinon $%
b_{i\sigma }$, which carry charge and spin degrees of freedom, respectively.
They are given by
\begin{align}
\mathcal{L}_{h}& =h_{i}^{\dagger }(d_{0}-iA_{0}^{s}-iA_{0}^{e}+\lambda
^{h})h_{i}-t_{h}\sum_{\alpha }\left( e^{iA_{\alpha }^{s}+iA_{\alpha
}^{e}}h_{i}^{\dagger }h_{i-{\hat{\alpha}}}+\mathrm{h.c.}\right) +\frac{u_{1}%
}{2}\left( h_{i}^{\dagger }h_{i}\right) ^{2},  \label{Lh} \\
\mathcal{L}_{s}& =\sum_{\sigma }b_{i\sigma }^{\dagger }\left( d_{0}-i\sigma
A_{0}^{h}+\lambda ^{s}\right) \,b_{i\sigma }-J_{s}\sum_{{\alpha },\sigma
}\left( e^{i\sigma A_{\alpha }^{h}}b_{i+{\hat{\alpha}}\sigma }^{\dagger
}b_{i-\sigma }^{\dagger }+\text{\textrm{h.c.}}\right) +\frac{u_{2}}{2}\left(
\sum_{\sigma }b_{i\sigma }^{\dagger }b_{i\sigma }\right) ^{2}.  \label{Ls}
\end{align}%
Equations (\ref{Lh}) and (\ref{Ls}) show that the bosonic holon and spinon
fields minimally couple to two lattice U(1) gauge fields, $A_{\mu }^{s}$ and
$A_{\mu }^{h}$, respectively, in 2+1 dimensions ($\mu =$ $\alpha ,0$, with $%
\alpha =x$, $y$). ($A_{\mu }^{e}$ is the external electromagnetic field
which solely couples to the holons\cite{Weng07}.) The parameters $t_{h}$ and
$J_{s}$ are effective hopping integral and superexchange coupling,
respectively, and $d_{0}$ is the difference operator defined along the
imaginary time axis. The Lagrangian multipliers, $\lambda ^{h,s}$, control
the total numbers of holons and spinons, and the last terms in $\mathcal{L}%
_{h,s}$ describe the on-site repulsions, which soften the hard-core boson
condition. It is noted that this effective theory is underpinned by a
so-called bosonic RVB order parameter: $\Delta ^{s}\equiv \sum_{\sigma
}\langle e^{i\sigma A_{\alpha }^{h}}b_{i+{\hat{\alpha}}\sigma }^{\dagger
}b_{i-\sigma }^{\dagger }\rangle \neq 0$, which describes a short-ranged
spin liquid state in the spinion background in the underdoped regime of a
doped Mott insulator\cite{Weng07}.

The gauge fields, $A_{\mu }^{h}$ and $A_{\mu }^{s}$, capture the essential
physics of the phase string effect\cite{Weng07} in a doped Mott insulator
via the following lattice mutual Chern-Simons term in Eq. (\ref{S})
\begin{equation}
\mathcal{L}_{\mathrm{MCS}}=\frac{i}{\pi }\, \epsilon ^{\mu \nu \lambda
}\left( A_{{\mu }}^{s}-2\pi \mathscr{N}_{\mu }^{s}\right) d_{\nu }\left( A_{{%
\lambda }}^{h}-2\pi \mathscr{N}_{\lambda }^{h}\right) \,,  \label{MCS}
\end{equation}%
with $\{ \mathscr{N}^{s,h}\}$ the integer fields due to the compactness of
the lattice field theory\cite{Ye1,Ye2}, which are important in restoring the
spin \textrm{SU(2) }symmetry. Without $\{ \mathscr{N}^{s,h}\}$, Eq. (\ref{LMCS}) is recovered the present
theory reduces to the original mutual Chern-Simons gauge theory formulation
of the phase string theory in Ref. \onlinecite{Kou05}.

\subsection{Low-energy effective theory for superconducting phase}

Superconductivity can naturally arise in a saddle-point state of the above
mutual Chern-Simons gauge theory, in which holons become Bose-condensed and
spinons are gapped\cite{Kou05,Ye1,Ye2}. In the folllowing, we shall derive a
general effective theory governing the low-energy fluctuations of the gauge
fields $A^{h,s}$ around this saddle-point. This new effective theory will
replace the conventional Ginzburg-Landau theory to describe the generalized
rigidity of   a non-BCS superconducting state.

By minimizing the action $S$ in Eq. (\ref{partition}) with regard to $%
A_{0}^{s}$ and $A_{0}^{h}$, one can find the saddle-point solution of $%
A^{h,s}$, i.e., $\bar{A}^{h,s}$ defined by
\begin{equation}
\epsilon ^{\alpha \beta }d_{\alpha }\bar{A}_{\beta }^{h}=\pi \delta ,\text{
\  \  \ }\bar{A}_{\alpha }^{s}=0,  \label{SP}
\end{equation}%
in the presence of a uniform distribution of the holon condensate with $\delta $ as the holon concentration. One can always choose the temporal components $\bar{A}_{0}^{s,h}=0$ by shifting $\lambda ^{h,s}$ in Eqs. (\ref{Lh}) and (\ref{Ls}). At the saddle point, one has $\overline{\mathscr{N}}^{s,h}=0$. Equation (\ref{SP}) can be understood as that a fictitious
uniform flux $\pi \delta $ is generated by the average distribution of the
holon condensate via $\bar{A}^{h}$, while the vanishing net flux in $\bar{A}^{s}$ is due to the fact that the spinons are short-range paired with $
\left
\langle \sum_{\sigma }\sigma b_{i\sigma }^{\dagger }b_{i\sigma
}\right
\rangle =0$ (where $\left \langle ...\right \rangle $ denotes the
expectation value at the saddle-point).

Around the saddle-point, the gauge field $A_{\mu }^{h}$ may be rewritten as
\begin{equation}
A_{\mu }^{h}=\bar{A}_{\mu }^{h}+a_{\mu }^{h}\, \text{.}  \label{amu}
\end{equation}%
Then an effective low-energy Lagrangian of the \textquotedblleft
weak\textquotedblright \ fluctuations of $A_{\mu }^{s}$ and $A_{\mu }^{h}$
around the saddle-point can be derived as follows

\begin{equation}
\mathcal{L}_{\mathrm{eff}}[A^{s},a^{h},A^{e}]=\frac{g_{\mu }^{h}}{2\pi ^{2}}%
(A_{\mu }^{s}+A_{\mu }^{e})^{2}+\frac{g_{\mu }^{s}}{2\pi ^{2}}\left(
\epsilon ^{\mu \nu \lambda }\partial _{\nu }a_{\lambda }^{h}\right) ^{2}+%
\frac{i}{\pi }\epsilon ^{\mu \nu \lambda }A_{\mu }^{s}\partial _{\nu
}a_{\lambda }^{h}\text{ \ .}  \label{effectiveL}
\end{equation}%
Here the first term on the right-hand-side (rhs) of Eq. (\ref{effectiveL})
arises from $\mathcal{L}_{h}$ in Eq. (\ref{Lh}) after considering the holon
condensation\cite{Ye2}, with $g_{0}^{h}=\pi ^{2}/u_{1}$ and $%
g_{1}^{h}=g_{2}^{h}=2\pi ^{2}\bar{\rho}_{h}t_{h}$. Note that the gauge field
$A^{s}$ is fixed in the unitary gauge by absorbing Goldstone boson
field $\partial _{\mu }\theta $ in the holon superfluid $h=\sqrt{\rho _{h}}%
e^{i\theta }$, and the superfluid density fluctuation $\delta \rho _{h}=\rho
_{h}-\bar{\rho}_{h}$ has been integrated out around $\bar{\rho}_{h}=\delta $.

The second term on the rhs of Eqs. (\ref{effectiveL}) originates from the
spinon term $\mathcal{L}_{s}$ in Eq. (\ref{Ls}) after integrating out the
spinon field $b$, with $g_{0}^{s}=\frac{\gamma }{m_{s}}$, $%
g_{1}^{s}=g_{2}^{s}=\frac{\kappa }{m_{s}}$ (the details are outlined in
Appendix \ref{appendix_s}). Here, $m_{s}$ denotes the spin mass gap which is
an important parameter as discussed in Appendix A below Eq. (\ref{Ls-1}).
The parameters $\gamma >0$ and $\kappa >0$ as given in Eq. (\ref{g}). Note
that in Refs. \onlinecite{Ye1,Ye2}, by assuming that the state is deep
inside the superconducting phase, the limit of $m_{s}\rightarrow \infty $
has been taken such that $g_{\mu }^{s}=0$ there. Equation (\ref{effectiveL})
corresponds to the general case for a finite spin mass gap $m_{s}$.

The last term on the rhs of Eq. (\ref{effectiveL}) is the mutual
Chern-Simons term with the high-energy fluctuations of $\mathscr{N}^{s,h}$
omitted. Note that the role played by $\mathscr{N}^{s}$ in constructing a
topological object like the magnetic flux quantization has been discussed in
Ref. \onlinecite{Ye2} and the references therein.

\subsection{Meissner effect}

To see that the effective Lagrangian $\mathcal{L}_{\mathrm{eff}}$ in Eq. (%
\ref{effectiveL}) indeed describes a superconducting state, one may further
integrate out $A^{s}$ and $a^{h}$ to arrive at a low-energy action
\begin{equation}
S_{\mathrm{eff}}[A^{e}]=\int d^{3}x\frac{M_{p}{}^{2}}{2}({\mathbf{A}}%
^{e\perp })^{2},  \label{Seff}
\end{equation}%
where the summation $\sum_{x}$ is replaced by integration in the continuum
limit, with the external magnetic field ${\mathbf{A}}^{e\perp }$ fixed in
the Coulomb gauge ($\nabla \cdot {\mathbf{A}}^{e\perp }=0$). Here the
transverse photon mass $M_{p}$ is given by
\begin{equation}
M_{p}=\frac{1}{\lambda _{p}}  \label{Mp}
\end{equation}%
with the London penetration depth
\begin{equation}
\lambda _{p}=\sqrt{\lambda _{0}{}^{2}+\kappa m_{s}^{-1}}\,  \label{lambda-p}
\end{equation}%
in which $\lambda _{0}\equiv \frac{1}{\sqrt{2t_{h}\bar{\rho}_{h}}}$.

Therefore, so long as $m_{s}\neq 0$, the system does exhibit the Meissner
effect with a finite $\lambda _{p}$ in response to an external magnetic
field. The physical electric current $\mathbf{J}^{e}$ satisfies the London
equation:
\begin{equation}
\mathbf{J}^{e}\equiv \frac{\delta S_{\mathrm{eff}}[A^{e}]}{\delta \mathbf{A}%
^{e\perp }}=\frac{1}{\lambda _{p}^{2}}\mathbf{A}^{e\perp }\text{.}
\label{London}
\end{equation}%
Equations (\ref{Mp}) and (\ref{lambda-p}) will recover the previous result%
\cite{Ye1,Ye2} obtained in the \textquotedblleft ideal
superconductor\textquotedblright \ limit of $m_{s}\rightarrow \infty $,
where the London penetration depth $\lambda _{p}$ approaches the
\textquotedblleft shortest\textquotedblright \ $\lambda _{0}$. Namely, in
the large spin-gap limit, the superfluidity of a holon condensate is well
protected.

However, in the opposite limit of $m_{s}\rightarrow 0$, i.e., with the spin gap closing
up , one finds $\lambda _{p}\rightarrow \infty $ in Eq. (\ref%
{lambda-p}) such that the Meissner effect disappears. It implies that the
superconducting phase coherence can be destroyed by the gapless spin
excitations at zero temperature, even in the presence of a bare holon
condensate with a finite $\lambda _{0}$. This is a peculiar manifestation
that the neutral spin degrees of freedom can strongly influence the
superconducting condensate. The quantum phase transition of
superconductivity occurring here is quit different from the conventional
transition of BCS-type as in the latter a charge (superconducting) gap must
be closed up at the Fermi surface. Physically, with $m_{s}\rightarrow 0$,
the proliferation of neutral spinons will disorder the superconducting phase
coherence via the $\pi $-vortices bound to them in the mutual Chern-Simons
gauge theory. Such a superconducting phase transition can also occur at a finite
temperature at $T=$ $T_{c}\sim m_{s}$ which is   of non-BCS-type (see
Ref. \onlinecite{Mei10}).

Finally it is noted that $m_{s}=0$ with a spinon condensation is a
necessary, but not a sufficient condition for the emergence of a true AFM
ordering in the mutual Chern-Simons gauge theory. A so-called \emph{Bose
insulating phase} (BI) with $m_{s}=0$ but $\lambda _{0}\neq 0$ has been
proposed\cite{Ye1,Ye2} to appear as an intermediate phase between the
superconducting and AFM phases. In other words, the present superconducting
phase can be terminated at a critical point defined by $m_{s}\rightarrow 0$
before a true AFM order can set in with further reducing doping
concentration near the half-filling.

\section{Topological Characterization of Superconductivity}

The effective low-energy Lagrangian $\mathcal{L}_{\mathrm{eff}}$ in Eq. (\ref%
{effectiveL}) has replaced the conventional Ginzburg-Landau equation to
describe superconductivity in the mutual Chern-Simons theory. Now we examine
the following question: What is the essential physical quantity that
captures the nature of superconductivity, since the BCS order parameter (as
governed by the conventional Ginzburg-Landau equation) no longer explicitly
appears in the low-energy effective theory?

To be sure, the off diagonal long order (ODLRO) of superconductivity is
still described by the Cooper pairing order parameter in the phase string
theory of doped Mott insulators\cite{Weng07}. Nevertheless, since the Cooper
pairing amplitude can persist well beyond the superconducting phase, as
determined\cite{Weng07} by the bosonic RVB parameter $\Delta ^{s}\neq 0$ and
the holon condensation $\left \langle h\right \rangle \neq 0$, the low-lying
excitations and the phase transition of superconducting phase are
essentially characterized by the topological phase fluctuations via the
mutual Chern-Simons gauge fields, $A^{s}$ and $a^{h}$, in the low-energy
effective Lagrangian (\ref{effectiveL}).

Physically, $A^{s}$ and $A^{h}$ are related to the spin and hole currents in
the mutual Chern-Simon gauge theory by the following equations of motion\cite%
{Kou05,Ye2}
\begin{equation}
j_{\alpha }^{s}=\frac{1}{\pi }\epsilon ^{0\alpha \beta }E_{\beta }^{s},\text{
\ }j_{\alpha }^{h}=\frac{1}{\pi }\epsilon ^{0\alpha \beta }E_{\beta }^{h},
\label{current}
\end{equation}%
where the internal \textquotedblleft electric\textquotedblright \ fields in
the imaginary time representation $E_{\alpha }^{s,h}\equiv -i\epsilon
^{\alpha \mu \nu }d_{\mu }A_{\nu }^{s,h}$ and the spin and charge currents
are defined as $j^{s/h}\equiv \frac{\delta {\mathcal{L}}_{s/h}}{\delta
A^{h/s}}$. Hence the low-lying fluctuations of $A^{s}$ and $a^{h}$ in Eq. (%
\ref{effectiveL}) actually describe the current fluctuations of the spin and
charge degrees of freedom, while the amplitude fluctuations of the matter
fields remain gapped in the superconducting phase.

\subsection{Wilson loops}

In order to characterize the gauge fluctuations of $A_{\mu }^{s,h}$, a pair
of Wilson loops have been introduced\cite{Ye1,Ye2}:
\begin{equation}
W^{s,h}[\mathcal{C}]\equiv \left \langle \hat{W}^{s,h}[\mathcal{C}]\right
\rangle ,  \label{Wilson}
\end{equation}%
where%
\begin{equation*}
\left \langle ...\right \rangle \equiv \mathcal{Z}^{-1}\int
D[A^{s},A^{h}]e^{-\sum_{x}\mathcal{L}_{eff}}...
\end{equation*}%
and
\begin{equation}
\hat{W}^{s,h}[\mathcal{C}]\equiv \exp \big [i\sum_{x}A_{\mu }^{s,h}(x)J_{%
\mathcal{C}}^{\mu }(x)\big],\, \,  \label{W-operator}
\end{equation}%
in which the (2+1)-dimensional unit current, for a test holon or a test
spinon, $J_{\mathcal{C}}^{\mu }(x)=+1$ $(-1)$ for the link $x\rightarrow x+%
\hat{\mu}$ $(x+\hat{\mu}\rightarrow x)$, and is zero otherwise on a close
loop $\mathcal{C}$.

Physically, the Wilson loop $W^{s}[\mathcal{C}]$ or $W^{h}[\mathcal{C}]$
probes the interaction of a pair of test holons or spinons at a distance $R$%
, defined by $V^{h}(R)$ or $V^{s}(R)$, via\cite{Smit}
\begin{equation}
V^{h,s}(R)\equiv -\lim_{T\rightarrow \infty }\frac{1}{T}\ln W^{s,h}[\mathcal{%
C}],  \label{confineV}
\end{equation}%
if $\mathcal{C}$ is taken as a spacetime rectangle with length $T$ ($R$) in
the imaginary time (spatial) direction. In Refs. \onlinecite{Ye1,Ye2}, in
the limit of large spin gap ($m_{s}\rightarrow \infty $), one finds $%
V^{h}(R)\sim \frac{\lambda _{0}^{2}}{2}$ and $V^{s}(R)\sim $ $\frac{\pi }{%
2\lambda _{0}^{2}}\, \ln R$. Namely the holons are deconfined while the
spinons are logarithmically confined. It is consistent with the present
saddle-point state of holon condensation discussed in Sec. II.

In the presence of a finite spin gap $m_{s}$, we find that the above
conclusion remains true, with the the confining potential, between a test
spinon and test anti-spinon, modified to
\begin{equation}
{V}^{s}(R)=\frac{\pi }{2\lambda _{p}^{2}}\ln \left( \frac{R}{R^{\ast }}%
\right) \,  \label{Vs}
\end{equation}%
where $\lambda _{0}$ is replaced by $\lambda _{p}$ defined in Eq. (\ref%
{lambda-p}) ($R^{\ast }$ is an ultraviolet cutoff of the distance). The
result indicates that the spinons will experience a
confinement-deconfinement transition at $\lambda _{p}\rightarrow \infty $ or
$m_{s}\rightarrow 0$, coinciding with the superconducting phase transition
critical point at $T=0$. On the other hand, $V^{h}(R)$ for holons remains
deconfined.

\subsection{ Topological characterization}

The Wilson loops, $W^{s,h}[\mathcal{C}]$, can provide important information
on the confinement/deconfinement of spinons and holons in the
superconducting phase as discussed above. However, to understand why $%
W^{s,h}[\mathcal{C}]$ behave as such, the topological nature of the mutual
Chern-Simons effective theory in Eq. (\ref{effectiveL}) has to be further
revealed.

For this purpose we define
\begin{equation}
e^{i\mathcal{I}}\equiv \mathcal{K}^{-1}\, \left \langle \hat{W}^{s}[\mathcal{%
C}_{1}]\, \hat{W}^{h}[\mathcal{C}_{2}]\right \rangle \,,  \label{topo}
\end{equation}%
with the normalization factor
\begin{equation}
\mathcal{K}\equiv W^{h}[\mathcal{C}_{1}]\,W^{s}[\mathcal{C}_{2}].  \label{K}
\end{equation}

\begin{figure}[h]
\centering
\includegraphics[width=10cm]{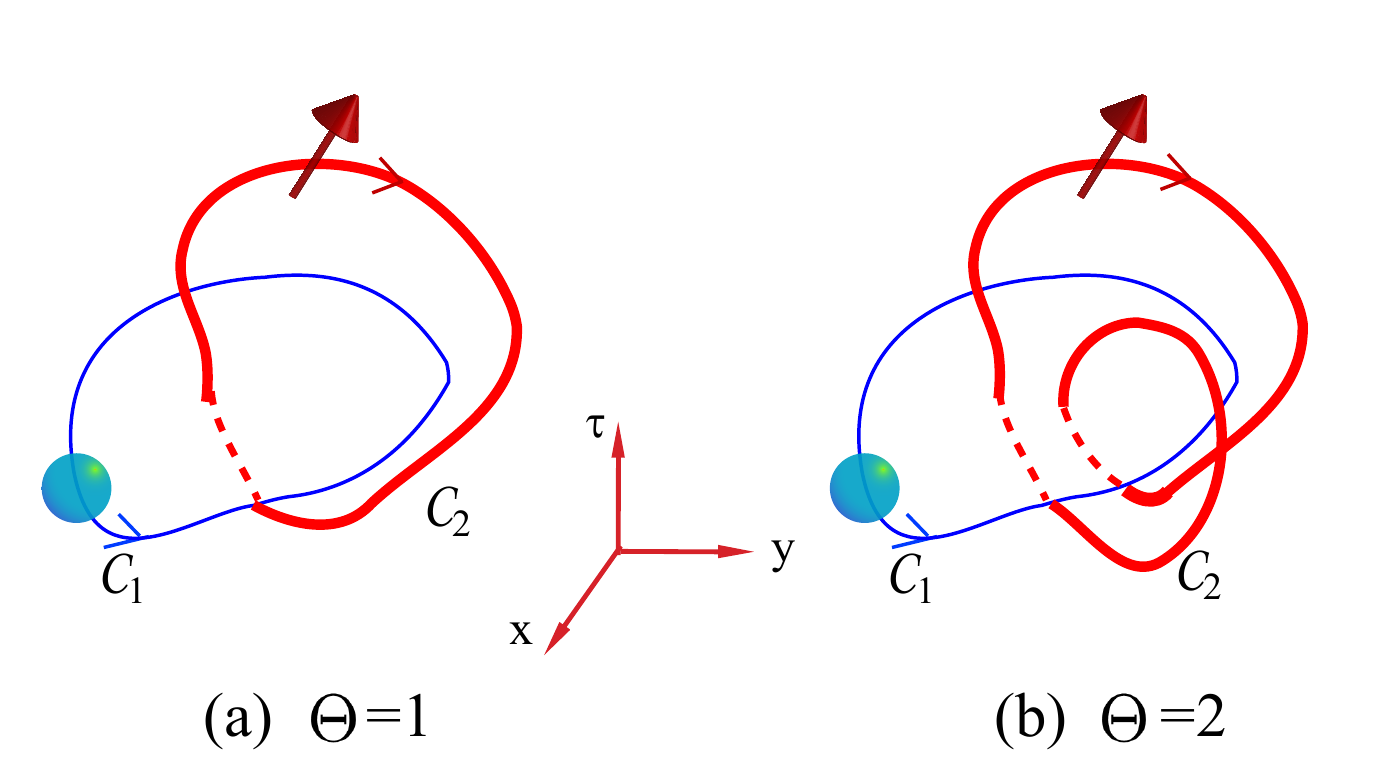}
\caption{(Color online). Schematic picture of two loops $\mathcal{C}_{1}$
and $\mathcal{C}_{2}$ in (2+1)-dimensional Euclidean spacetime. $\mathcal{C}%
_{1}$ is an arbitrary loop but restricted inside spatial plane (i.e., the xy
plane) at some imaginary time. $\mathcal{C}_{2}$ is an arbitrary loop with a
definite Gaussian linking number $\Theta \in \mathbb{Z}$ ($\Theta =1$ in
(a); $\Theta =2$ in (b)) that characterizes the number of times that $%
\mathcal{C}_{2}$ winds around $\mathcal{C}_{1}$. The ball (in blue) and the
arrow (in red) stand for the test holon and test spinon, respectively. The
arrows on the two loops indicate the direction of test spinon current and
test holon current.}
\label{figure_geometry}
\end{figure}

Let us first consider the case of a \emph{pure }mutual Chern-Simons term (%
\ref{MCS}) without coupling to the matter fields [i.e., omitting the first
two terms in Eq. (\ref{effectiveL})], a straightforward calculation shows
\begin{equation}
\mathcal{I}=\pi \Theta  \label{link number}
\end{equation}%
where $\Theta =0,\pm 1,\pm 2,\cdots $, denotes the Gaussian linking number,
which is the winding number of the close loops, $\mathcal{C}_{1}$ and $%
\mathcal{C}_{2},$ as shown in Fig. \ref{figure_geometry} (a) and (b). To
obtain Eq. (\ref{link number}), we write down the explicit expression of $%
\mathcal{I}$ in a pure mutual Chern-Simons gauge theory,
\begin{equation*}
e^{i\mathcal{I}}=\mathcal{K}^{-1}\int D[A^{s},a^{h}]e^{-\int d^{3}x\left(
\frac{i}{\pi }\epsilon ^{\mu \nu \lambda }A_{\mu }^{s}\partial _{\nu
}a_{\lambda }^{h}-A_{\mu }^{s}J_{\mu }^{\mathcal{C}_{1}}-a_{\mu }^{h}J_{\mu
}^{\mathcal{C}_{2}}\right) }
\end{equation*}%
where, the unit spacetime currents $J^{\mathcal{C}_{1}}$ and $J^{\mathcal{C}%
_{2}}$ keep track of spacetime loops, $\mathcal{C}_{1}$ and $\mathcal{C}_{2}$%
, respectively. Integrating over $A^{s}$ leads to a flux constraint on $%
a^{h} $ configuration, namely, $\epsilon ^{\mu \nu \lambda }\partial _{\nu
}a_{\lambda }^{h}=\pi J_{\mu }^{C_{1}}$. It means that the gauge field $a^{h}$
forms a spacetime closed $\pi$-fluxtube along $\mathcal{C}_{1}$. $\int
d^{3}xa_{\mu }^{h}J_{\mu }^{\mathcal{C}_{2}}=\oint_{\mathcal{C}_{2}}a_{\mu
}^{h}dx_{\mu }$ denotes the spacetime flux piercing $\mathcal{C}_{2}$.
Noting that there are $\Theta $ times for the fluxtube $\mathcal{C}_{1}$
piercing $\mathcal{C}_{2}$, one arrives at $e^{i\mathcal{I}}=e^{i\pi \Theta
} $, with $\mathcal{K}=1$. Ignoring the $2\pi $-periodicity, we thus obtain
Eq. (\ref{link number}). In Witten's seminal paper\cite{Witten89}, a similar
Gaussian linking number structure was found in the Chern-Simons gauge
theory, which is related to the Jones polynomial in the Knot theory.

\begin{figure}[h]
\centering
\includegraphics[width=10cm]{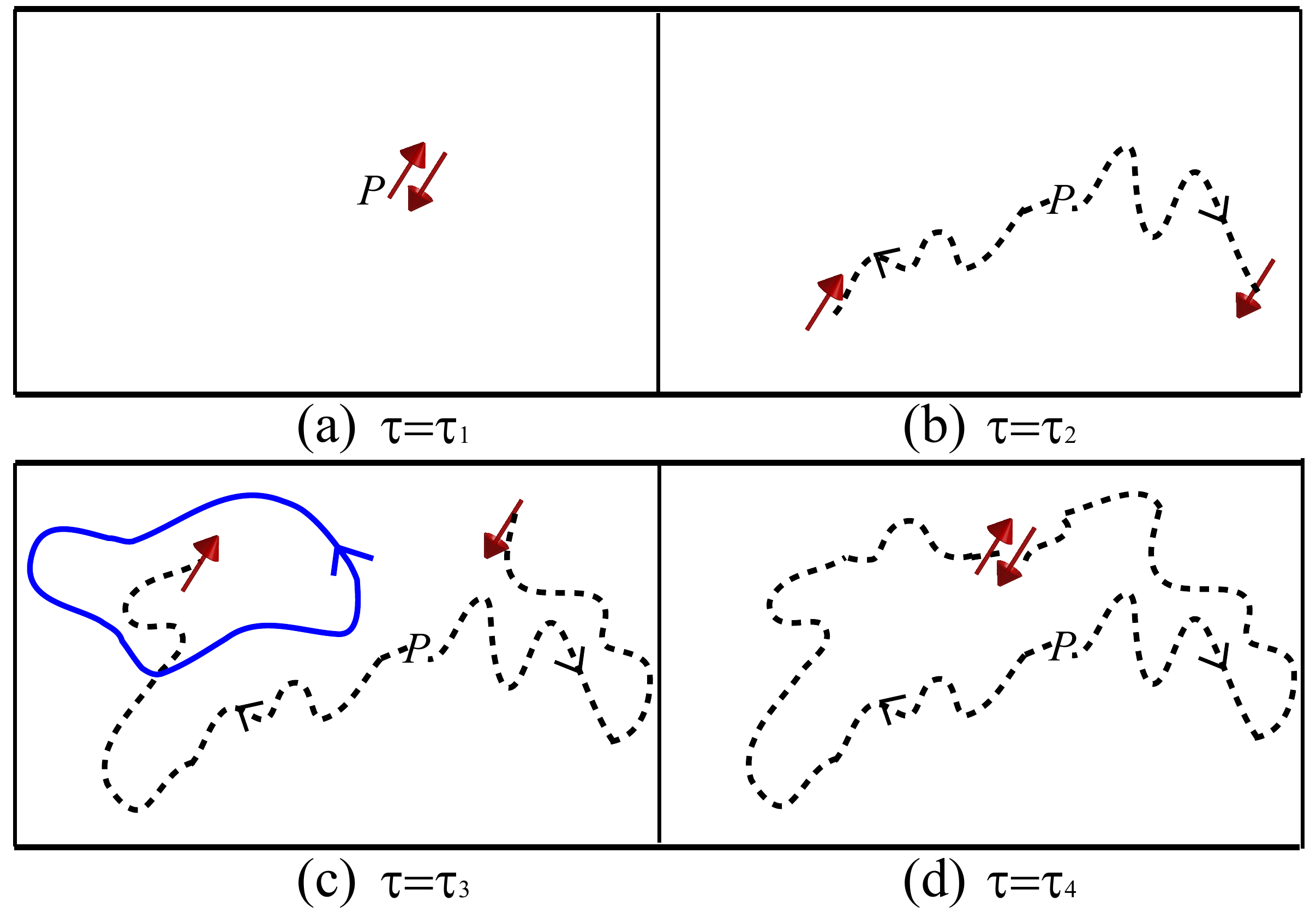}
\caption{(Color online). Gedankenexperiment (Gaussian linking number $g=1$
for simplicity). (a-d) are four snapshots of spatial plane subsequently at
imaginary time $\protect \tau _{1}<\protect \tau _{2}<\protect \tau _{3}<%
\protect \tau _{4}$. (a) A pair of test spinons with opposite spin indices
are created at a spatial site \textquotedblleft $P$\textquotedblright \ at $%
\protect \tau _{1}$. (b) Two spinons arrive at new spatial positions at $%
\protect \tau _{2}$ via the dashed trajectories. (c) At $\protect \tau _{3}$,
a holon current circuit $\mathcal{C}_{1}$ is located in plane, indicating
that an Aharonov-Bohm phase around $\mathcal{C}_{1}$ is picked up by the
test holon. (d) At $\protect \tau _{4}$, the pair of test spinons meet each
other again at some position and then are annihilated into vacuum
simultaneously. }
\label{figure_experiment}
\end{figure}

In Fig. \ref{figure_experiment}, a gedankenexperiment is shown to illustrate
the case of winding number $\Theta =1$ in the (2+1)-dimensional Euclidean
spacetime, with a pair of test spinon and antispinon creating and
annihilating on a close loop $\mathcal{C}_{2}$. If the background spin
degrees of freedom are gapped with $m_{s}\longrightarrow \infty $, the test
spinons cannot induce spin excitations from the ground state and thus the
net contribution to $e^{i\mathcal{I}}$ comes from the $\pi $ flux phase
shift (with $\Theta =1)$ mentioned above due to the pure mutual Chern-Simons
action.

On the other hand, if the spin gap $m_{s}$ becomes finite, then the spin
excitations can be induced from the spinon background to renomalize $%
\mathcal{I}$. Now we consider the full $\mathcal{L}_{\mathrm{eff}}$ given in
Eq. (\ref{effectiveL}). By choosing the close loop $\mathcal{C}_{1}$ of the
test holon within the spatial $xy$-plane, one finally arrives at the
following compact result (for the details, see Appendix B):
\begin{equation}
\mathcal{I}=\pi \Theta \, \left( \lambda _{0}/{\lambda _{p}}\right) ^{2}.
\label{I}
\end{equation}%
We see that $\mathcal{I}$ remains to be real and positive as a topological
quantity, which is invariant under a continuous deformation of loops $%
\mathcal{C}_{1}$ and $\mathcal{C}_{2}$ for a given Gaussian linking number $%
\Theta $, as illustrated in Fig. \ref{figure_geometry}. But $\mathcal{I}$ is
renormalized by a factor $\left( \lambda _{0}/{\lambda _{p}}\right) ^{2}$
which is solely determined by the spin gap $m_{s}$. In particular, in the
limit of $m_{s}\rightarrow 0$, one finds $\mathcal{I}\rightarrow 0$.

If one rewrites Eq. (\ref{I}) by
\begin{equation}
\lambda _{p}=\lambda _{0}\cdot \sqrt{{\pi }/{\mathcal{I}}}\text{,}
\label{lambda-p-1}
\end{equation}%
where $\lambda _{p}$ becomes a function of topological order parameter $%
\mathcal{I}$ for a given Gaussian linking number $\Theta =1$, then the limit
$\lambda _{p}\rightarrow \infty $, i.e., the disappearance of
superconductivity, will correspond to $\mathcal{I}=0$. In other words, the
topological quantity $\mathcal{I}$ can be regarded as the superconducting
order parameter in the effective theory of Eq. (\ref{effectiveL}), and the
quantum phase transition at $T=0$ occurs when $\mathcal{I}$ gets fully
screened by the gapless spin excitations at $m_{s}=0$.

As illustrated in Fig. \ref{figure_geometry}, $\mathcal{I}\neq 0$ indicates
that a test spinon can be always \textquotedblleft felt\textquotedblright \
by a vortex-like current loop of a test holon if their mutual winding number
$\Theta \neq 0$. Physically, since the holons are in the Bose condensed
state, a spinon excitation will then induce a supercurrent vortex response
from the condensed holons throughout the sample to form a so-called
spinon-vortex\cite{Qi07}. This is the origin of spinon confinement discussed
above, since each spinon-vortex composite is logarithmically divergent in
energy which will result in vortex-antivortex confinement. In a
non-superconducting phase where the spinons become deconfined due to the
vortex-antivortex disassociation and proliferation, one finds\ $\mathcal{I}%
=0 $ in the gedankenexperiment shown in Fig. \ref{figure_experiment}.
Namely, $\mathcal{I}=0$ describes a phase disordered superconductor.

\subsection{Experimental prediction}

\begin{figure}[h]
\centering
\includegraphics[width=8cm]{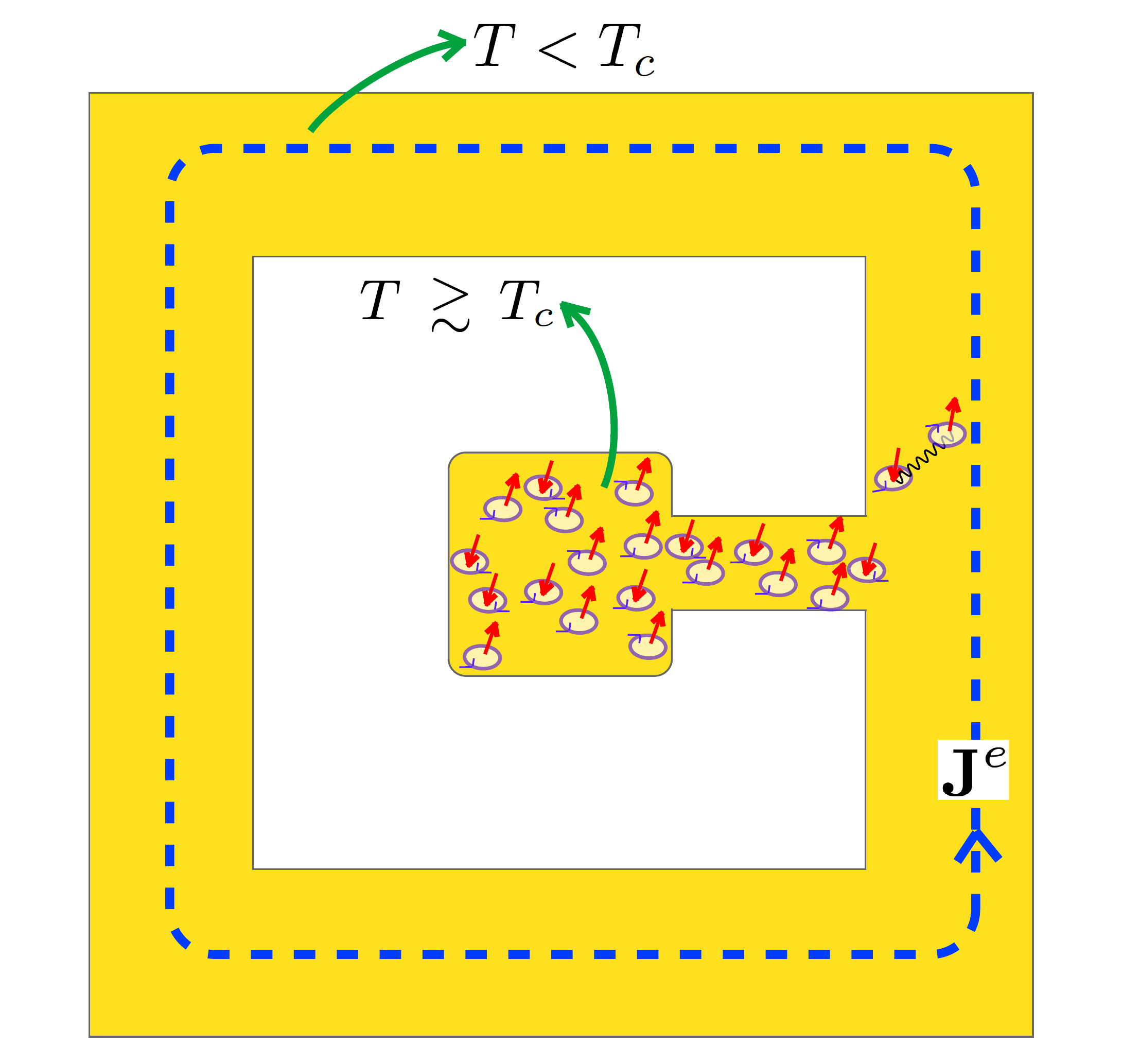}
\caption{(Color online) A proposed experiment to detect mutual statistics in
a mutual Chern-Simons superconductor. The temperature in the middle semi-island is raised
slightly above the superconducting critical temperature $T_{c}$, while the
outside part remains below $T_{c}$ and spinons are confined into vortex-antivortex pairs. The confinement is indicated by the black wavy line between two spinons.
The small directional discs (red arrows) represent vortices (spinons). In such a mutual Chern-Simons superconductor at $T\gtrsim $ $%
T_{c}$, spinons become deconfined to form free spinon-vortices \protect \cite%
{Qi07}. Then, the flow of such free spinons from the semi-island towards the
cooler end connecting to the outside superconducting ring is expected to
generate random supercurrents in the latter as indicated by $\mathbf{J}^{e}$ (blue dashed line). }
\label{figure_prediction}
\end{figure}

We propose an experimental setup to probe the topological characterization $%
\mathcal{I}$ of the superconducting phase in Fig. \ref{figure_prediction}.
Note that $\mathcal{I}\neq 0$ reflects the mutual statistical interaction
between spin and charge degrees of freedom (cf. Fig. \ref{figure_geometry}).
By using the idea illustrated in the gedankenexperiment of Fig. \ref%
{figure_experiment}, one may inject free \textquotedblleft
test\textquotedblright \ spinons by heating up the middle semi-island part
of the sample in Fig. \ref{figure_prediction} to a temperature slightly
higher than $T_{c}$, such that free spinons can be excited and flow towards
the lower $T$ end connecting to the outside superconducting ring. With $%
\mathcal{I}\neq 0$ inside the ring, one expects that such an injection of
free spinons can generate supercurrents in the superconducting ring in a
form of significant noise spikes due to the unpolarized spinon-vortex
effect.

Instead of raising the temperature, one may also apply an in-plance magnetic
field $B^{e}$ to reduce the spinon mass gap $m_{s}$ inside the semi-island
part by $\widetilde{m_{s}}=m_{s}-\frac{1}{2}g_{L}\mu _{B}B^{e}$, where $%
g_{L} $ is the Land\'{e} factor and $\mu _{B}$ is the Bohr magneton. Then a
sufficiently strong $B^{e}$ may drive the\ semi-island part into a
spinon-vortex liquid state with $\mathcal{I}=0$ via the Zeeman effect\cite%
{Qi07}. At the same time, the gradient of $B^{e}$ can also naturally cause
the free spinons flow into the superconducting ring to induce the similar
supercurrent response shown in Fig. \ref{figure_prediction}.

\subsection{Emergence of fermionic excitations: A proof-of-principle
derivation}

The mutual Chern-Simons theory given in Eq. (\ref{S}) is an all-boson
description of doped Mott insulators. In the above subsections we have
studied the low-lying collective fluctuations of this theory based on the
effective Lagrangian (\ref{effectiveL}). An important question that arises
naturally is if and how a low-energy fermionic excitation can \emph{emerge}
in such an all-boson theory, which is independent of the above collective
excitations.

To give a proof of principle, we first note that the mutual Chern-Simons
term in Eq. (\ref{effectiveL}) can be replaced by an auxiliary Lagrangian in
real-time representation:
\begin{equation}
\mathcal{L}_{\text{\textrm{aux}}}=\sum_{f=1,2}\big(\overline{\Psi }_{f}i%
\slashed{\mathcal{D}}\Psi _{f}-\Delta \overline{\Psi }_{f}\Psi _{f}\big)\,,
\label{aux}
\end{equation}%
where $\Psi _{f}$ is an auxiliary 4-component Dirac spinor in
(2+1)-dimensional spacetime with \textquotedblleft flavor\textquotedblright
\ indices $f=1,2$. $\slashed{\mathcal{D}}\equiv \gamma ^{\mu }\mathcal{D}%
_{\mu }$. Here $\gamma $-matrices are defined in 4-dimensional reducible
representation of 2-dimensional Clifford algebra, i.e., $\gamma
^{0}=diag(\sigma ^{3},-\sigma ^{3})\,,\gamma ^{\alpha }=diag(\sigma ^{\alpha
},-\sigma ^{\alpha })$, where $\alpha =1,2$ and $\sigma ^{1,2,3}$ are Pauli
matrices. The covariant derivative $\mathcal{D}_{\mu }$ is defined as $%
\mathcal{D}_{\mu }\equiv \partial _{\mu }+iA_{\mu }^{s}+i\tau _{3}a_{\mu
}^{h}\,$ The $4\times 4$ matrix $\tau _{3}=diag(\mathbb{I},-\mathbb{I})$,
where $\mathbb{I}$ is the $2\times 2$ unit matrix. The second term in Eq. (%
\ref{aux}) is mass term with $\Delta >0$.

\begin{figure}[h]
\centering
\includegraphics[width=8cm]{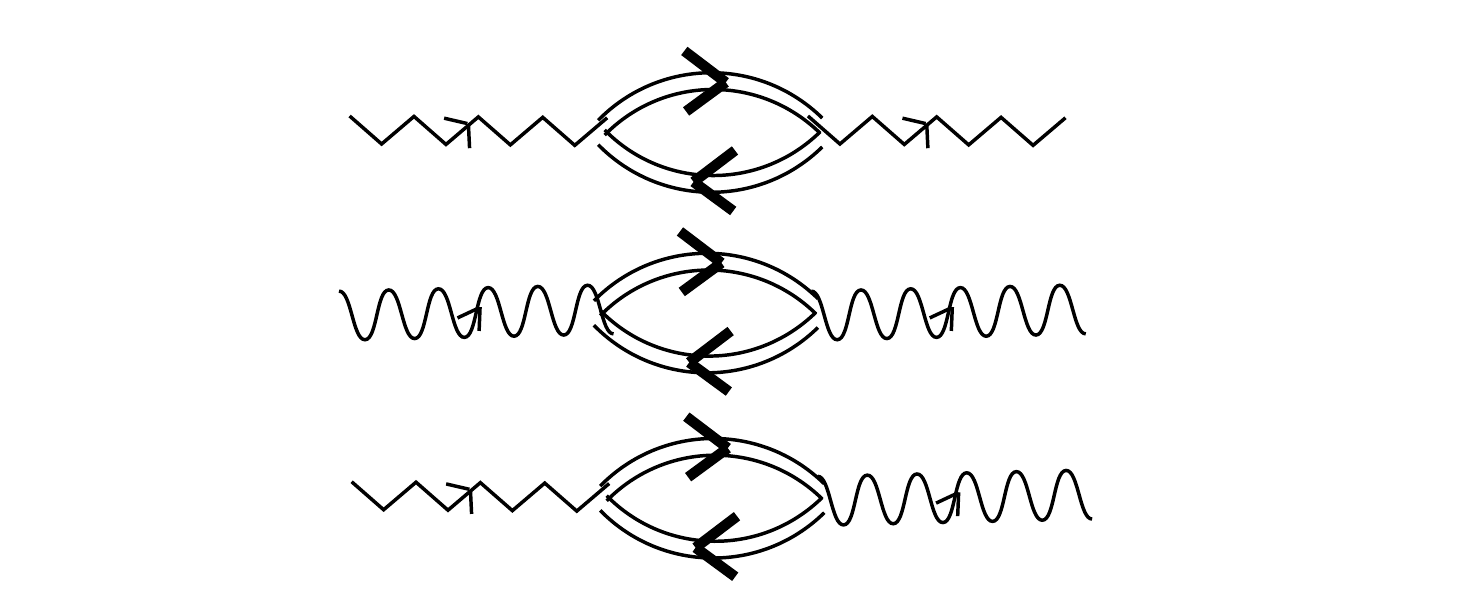}
\caption{One-loop quantum correction. Zigzag, wavy, and double lines
represent $A^{s}$, $a^{h}$, and $\Psi $, correspondingly. The first two
diagrams correspond to Maxwell terms, while the last one gives the
non-vanishing mutual Chern-Simons vertex.}
\label{vertex}
\end{figure}

$\mathcal{L}_{\text{\textrm{aux}}}$ preserves a $U(1)_{A}\times U(1)_{V}$
gauge symmetry where subscripts $A$ and $V$ denote axial vector type and
ordinary vector type $U(1)$ gauge invariance, respectively. It originates
from the fact that $A_{\mu }^{s}$ is an ordinary vector field with gauge
group generator \textquotedblleft 1\textquotedblright \ while $a^{h}$ is an
axial vector field with gauge group generator $\tau _{3}$ \cite{Kou05}. The
parity and time-reversal symmetries are conserved even in the presence of
finite mass gap which may be expressed as: $\Delta \overline{\Psi }_{f}{\Psi
}_{f}=\Delta \Psi _{f}^{\dagger }\gamma _{0}\Psi _{f}=\Delta \Psi
_{f+}^{\dagger }\sigma ^{3}\Psi _{f+}-\Delta \Psi _{f-}^{\dagger }\sigma
^{3}\Psi _{f-}$, where, $\Psi _{f+}$ ($\Psi _{f-}$) is upper (lower)
2-component spinor of $\Psi _{f}$. Here, $\pm $ denotes the chirality\cite%
{Vafa84}. In the case of parity ($\hat{\mathbb{P}}$ is parity operator),
\begin{equation}
\Psi _{f}\rightarrow \hat{\mathbb{P}}\Psi _{f}=\left(
\begin{array}{cc}
0 & \sigma ^{1} \\
\sigma ^{1} & 0%
\end{array}%
\right) \left(
\begin{array}{c}
\Psi _{f+} \\
\Psi _{f-}%
\end{array}%
\right) =\left(
\begin{array}{c}
\sigma ^{1}\Psi _{f-} \\
\sigma ^{1}\Psi _{f+}%
\end{array}%
\right) \,.  \label{Psif}
\end{equation}%
It is easy to check that the mass term is invariant under such parity
transformation in (2+1)-dimensional spacetime. Integrating over $\Psi _{f}$
leads to  a pure mutual Chern-Simons term by ignoring Maxwell terms ($\propto $
momentum squared) in the one-loop quantum correction (Fig. \ref{vertex}) in
the long wavelength limit $|p|\ll \Delta $. The key steps of derivation are
given as follows.
As shown in Fig. \ref{vertex}, the vacuum polarization for $A^{s}$ and $%
a^{h} $ at one-loop level\cite{Pisarski84,Deser82} ($p$ is a 3-dimensional
Euclidean momentum):
\begin{equation*}
\Pi (p)=\frac{\mathcal{N}_{f}}{4\pi p^{2}}\bigg[2\Delta +\frac{p^{2}-4\Delta
^{2}}{p}\arcsin \bigg(\frac{p}{(p^{2}+4\Delta ^{2})^{1/2}}\bigg)\bigg].
\end{equation*}%
where the total flavor number $\mathcal{N}_{f}=2$. The expansion around
small $p$ is given by
\begin{equation*}
\Pi (p)=\frac{\mathcal{N}_{f}}{6\pi \Delta }-\frac{\mathcal{N}_{f}p^{2}}{%
60\pi \Delta ^{3}}+\frac{3\mathcal{N}_{f}p^{4}}{1120\pi \Delta ^{5}}+\cdots .
\end{equation*}%
The propagators of gauge fields can be constructed in the Landau gauge and
the Maxwell terms read out: $-\frac{(f_{\mu \nu }^{s})^{2}}{12\pi \Delta }-%
\frac{(f_{\mu \nu }^{h})^{2}}{12\pi \Delta }$ with $f_{\mu \nu }^{s}\equiv
\partial _{\mu }A_{\mu }^{s}-\partial _{\nu }A_{\mu }^{s}$ and $f_{\mu \nu
}^{h}\equiv \partial _{\mu }a_{\mu }^{h}-\partial _{\nu }a_{\mu }^{h}$.
Meanwhile, the nonvanishing mutual Chern-Simons term for $A^{s}$ and $a^{h}$
in Fig. \ref{vertex} is also radiatively generated without breaking parity
and time-reversal symmetries. Note that, upper(lower)-component spinor has
mass $\Delta $ ($-\Delta $):
\begin{equation*}
\Pi _{\mu \nu }^{\sigma }(p)=-\sigma \Delta \mathcal{N}_{f}\int \frac{d^{3}k%
}{(2\pi )^{3}}\frac{Tr(\widetilde{\gamma }_{\mu }\widetilde{\gamma }%
_{\lambda }p^{\lambda }\widetilde{\gamma }_{\nu })}{(k^{2}+\Delta
^{2})((p+k)^{2}+\Delta ^{2})}=\frac{\sigma \Delta \mathcal{N}_{f}}{4\pi }%
\epsilon _{\mu \nu \lambda }p^{\lambda }\int_{0}^{1}d\ell \frac{1}{\sqrt{%
\Delta ^{2}+\ell (1-\ell )p^{2}}}
\end{equation*}%
where, the superscript $\sigma =\uparrow (1),\downarrow (-1)$ denoting upper
and lower 2-component spinor contributions respectively, and $\widetilde{%
\gamma }_{\mu }=i\{ \sigma ^{1},\sigma ^{2},\sigma ^{3}\}$ with identity: $%
Tr(\widetilde{\gamma }_{\mu }\widetilde{\gamma }_{\nu }\widetilde{\gamma }%
_{\lambda })=2\epsilon _{\mu \nu \lambda }$. In the sufficiently low
momentum transfer limit ($p\rightarrow 0$), it is expanded as power series
of $p$:
\begin{align}
\Pi _{\mu \nu }^{\sigma }(p)=\frac{\sigma \mathcal{N}_{f}}{4\pi }\epsilon
_{\mu \nu \lambda }p^{\lambda }\bigg(1-\frac{p^{2}}{12\Delta ^{2}}+\frac{%
p^{4}}{80\Delta ^{4}}+\cdots \bigg)\,.\label{eq_sigma}
\end{align}
In Eq. (\ref{eq_sigma}), the sign factor \textquotedblleft $\sigma $%
\textquotedblright \ renders vanishing of Abelian Chern-Simons terms ($\sim
\epsilon ^{\mu \nu \lambda }A_{\mu }^{s}\partial _{\nu }A_{\lambda }^{s}$
and $\sim \epsilon ^{\mu \nu \lambda }a_{\mu }^{h}\partial _{\nu }a_{\lambda
}^{h}$). But due to the structure of matrix $\tau _{3}$, the crossing term,
i.e the mutual Chern-Simons term survives as expected. Mathematically, the
above quantum corrections are organized as below:
\begin{align}
\frac{1}{2}\sum_{\sigma =\pm }\left( A_{\mu }^{s}+\sigma a_{\mu }^{h}\right)
\frac{\sigma \mathcal{N}_{f}}{4\pi }\epsilon ^{\mu \nu \lambda }\partial
_{\nu }\left( A_{\lambda }^{s}+\sigma a_{\lambda }^{h}\right) =\frac{1}{\pi }%
\epsilon ^{\mu \nu \lambda }A_{\mu }^{s}\partial _{\nu }a_{\lambda }^{h}\,.
\end{align}%
where the Maxwell terms are ignored in the long wavelength limit.

\begin{figure}[h]
\centering
\includegraphics[width=8cm]{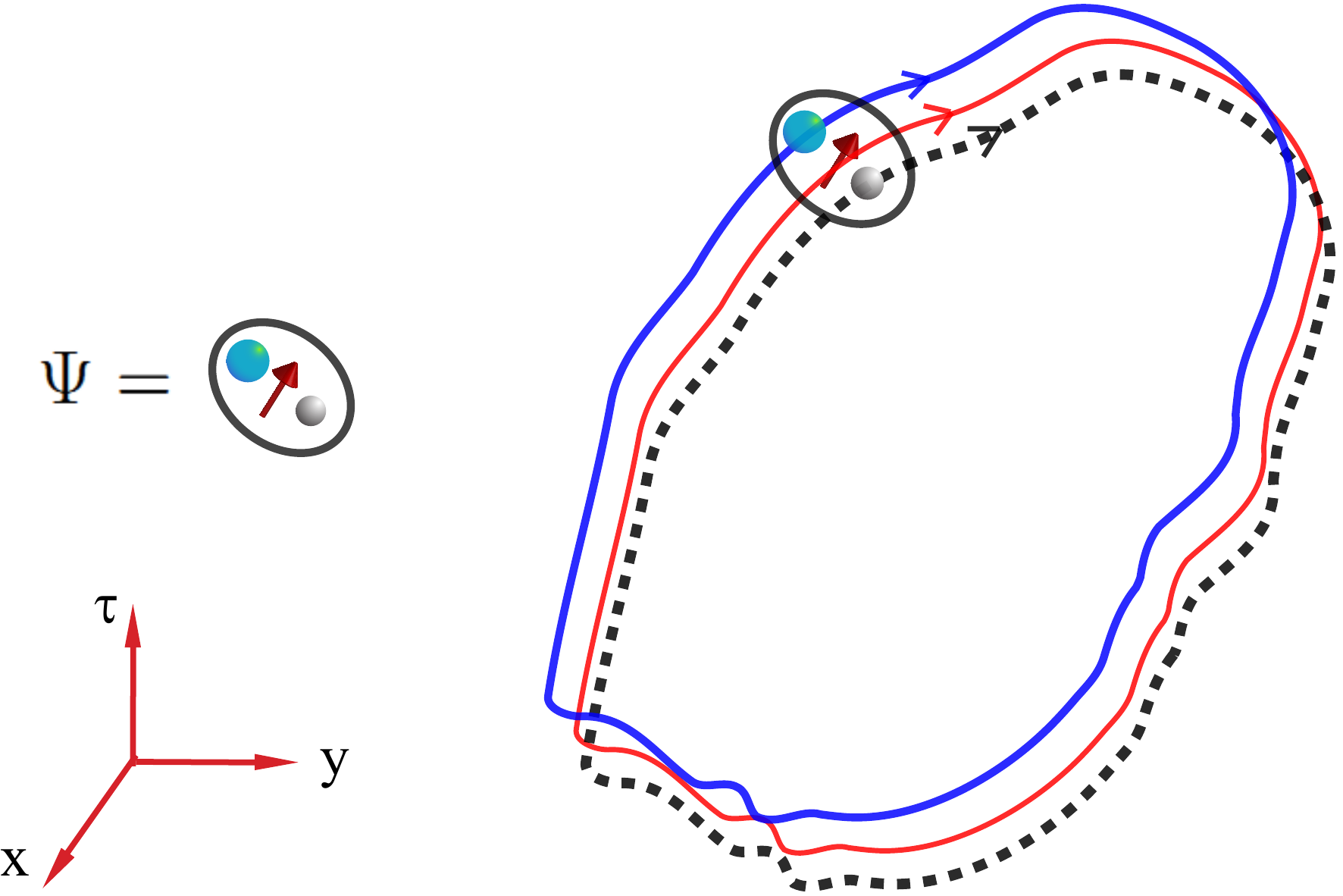}
\caption{(Color online). $\Psi $ is composed by a spinon (red arrow), a
holon (blue ball) and a Dirac fermion (small grey ball $\Psi _{f}$). Its
spacetime loop is composed by three loops rendering gauge charge neutrality.}
\label{qploop}
\end{figure}

It is important to note that the fermionic auxiliary field $\Psi _{f}$ is
not gauge-neutral as it still couples to $A_{\mu }^{s}$ and $a_{\mu }^{h}\,$%
in Eq. (\ref{aux}). A true fermion excitation of quasiparticle should be
gauge-neutral without knowing $A^{s}$ and $a^{h}$. It is only possible if
there exists a residual attractive potential such that a holon and a spinon
excited from the condensates can form a bound state properly with the
auxiliary field $\Psi _{f}$. The resulting object $\Psi $ will be both \emph{%
fermionic} and \emph{coherent} because all statistical gauge charges carried
by $\Psi _{f}$ can be exactly canceled by that of the spinon and holon
partners. In Fig. \ref{qploop}, the spacetime loop of a composite-particle $%
\Psi $ is illustrated as a gauge neutral and thus physically observable
elementary excitation at low energies.

In particular, it carries the right quantum number as a Bogoliubov
quasiparticle. As noted before, to produce the right prefactor $i/\pi $ of
the pure mutual Chern-Simons term, $\Psi _{f}$ has totally $2\times 2=4$
branches of positive energy solutions, which are consistent with the
existence of totally four quasiparticle nodal regions in the Brillouin zone,
although the detailed energy spectrum depends on the residual attractive
potential which goes beyond the present long-wavelength effective
description. On the other hand, the existence of a \emph{nodal} fermionic
Bogoliubov quasiparticle has been recently demonstrated based on a ground
state wavefunction approach to the $t$-$J$ model in the phase string
formulation, in which the quasiparticle as a similar bound state has been
explicitly shown\cite{Weng11}. Therefore, the present field-theory
formulation provides a \emph{proof-of-principle} for a fermionic
quasiparticle mode emerging in the mutual Chern-Simons theory.

%

\section{Conclusion}

In this work, we have obtained an effective theory of the low-lying
fluctuations in the superconducting state of the mutual Chern-Simons theory.
In contrast with a conventional Ginzburg-Landau action for a BCS
superconductor, the present effective description in Eq. (\ref{effectiveL})
deals with peculiar phase fluctuations in the doped Mott insulator via a
pair of mutual Chern-Simons gauge fields, which are related to the
fluctuations of charge and spin currents of the condensate.

The superconducting phase coherence is found to be properly characterized by
a topological order parameter which measures the mutual entanglement of the
charge and spin currents in the low-energy effective theory. In particular,
a zero-temperature quantum phase transition is shown to occur, due to the
vanishing of this topological order parameter, when the spin gap is reduced
to zero at low doping.

We have discussed a possible experimental probe into the long-range
entanglement between the supercurrents and the neutral spin excitations as a
unique signature of this non-BCS superconducting state. We have also given a
proof-of-principle how a fermionic quasiparticle can emerge in the mutual
Chern-Simons theory. Together with the previously identified flux
quantization\cite{Kou05,Ye1} at $hc/2e$, the non-BCS superconductivity in
the mutual Chern-Simons theory is well established.

\begin{acknowledgments}
We acknowledge stimulating discussions with Z. C. Gu, S.-P. Kou, Z.-X. Liu, C.-S. Tian, Y. Wang, Z. Wang, N. Xu, and Y.-Z. You. This work is supported by NSFC grant No. 10834003, by MOST National Program for Basic Research grant No. 2009CB929402, No. 2010CB923003.
\end{acknowledgments}

\appendix

\section{Path-integration over spinons}

\label{appendix_s}

First of all, we note that the spinon Lagrangian $\mathcal{L}_{s}$ in Eq. (%
\ref{Ls}) may be further simplified to a CP$(1)$ form in the
long-wavelength, low-energy regime as follows\cite{Kou05}
\begin{equation}
\mathcal{L}_{s}\simeq \sum_{\sigma }\frac{1}{2g}\bigg[\left \vert (\partial
_{\mu }-i\sigma {A}_{\mu }^{h})z_{\sigma }\right \vert
^{2}+m_{s}^{2}|z_{\sigma }|^{2}\bigg]\,,  \label{Ls-1}
\end{equation}%
Here the temporal component is rescaled by the spin wave velocity $%
c_{s}\equiv \sqrt{J_{s}(\lambda _{s}+4J_{s})}$ with the coupling constant $%
g=c_{s}/\left( 2J_{s}\right) $, and $m_{s}$ is treated as a dynamically
generated spinon mass enforcing $\sum_{\sigma }|z_{\sigma }|^{2}=1$.

The problem of studying the fluctuation of ${A}_{\mu }^{h}$, or ${a}_{\mu
}^{h}$, in Eq. (\ref{Ls-1}) is then transformed into calculating the vacuum
polarization tensor in (2+1)-dimensional scalar quantum electrodynamics in
the presence of a uniform background magnetic field of $\bar{A}^{h}$ defined
in Eq. (\ref{SP}). The technical challenge arises from the fact that because
of $\bar{A}^{h}$, the momentum is no longer a good quantum number.
Nevertheless the issue is resolved by using the so-called \textquotedblleft
string-inspired\textquotedblright \ formalism\cite{Schubert01}. Following
the procedures and conventions in Ref. \onlinecite{Schubert01}, we obtain
the polarization tensor:
\begin{equation}
\Pi _{\mu \nu }(k)=2\times \frac{-1}{(4\pi )^{\frac{D}{2}}}\, \times
\int_{0}^{\infty }\frac{dT}{T}T^{2-\frac{D}{2}}\,e^{-m_{s}^{2}T}\text{det}^{-%
\frac{1}{2}}\left[ \frac{\sin (Z)}{Z}\right] \int_{0}^{1}d\ell I^{\mu \nu }
\label{bubble}
\end{equation}%
where the prefactor $2$ comes from the summation over the spin index, $T$
denotes the Schwinger proper time, and spacetime dimension $D=3$. Here $Z$
is a $3\times 3$ matrix, defined by $Z=\pi FT$ with
\begin{equation}
F=\left(
\begin{array}{ccc}
0 & 0 & 0 \\
0 & 0 & \pi \, \delta \, \\
0 & -\pi \, \delta \, & 0%
\end{array}%
\right) \,,\qquad
\end{equation}%
The determinant in Eq. (\ref{bubble}) can be simply factorized into:
\begin{equation}
\text{det}^{-\frac{1}{2}}\left[ \frac{\sin (Z)}{Z}\right] =\frac{\pi \,
\delta \,T}{\sinh (\pi \, \delta \,T)}\text{,}
\end{equation}%
and the transversality of Lorentz tensor $I_{\mu \nu }$ is given by
\begin{equation}
I_{\mu \nu }=-(1-2\ell )\, \frac{\sinh (\pi \, \delta \,T(1-2\ell ))}{\sinh
(\pi \, \delta \,T)}\left( k_{\mu }k_{\mu }-\delta _{\mu \nu }k^{2}\right) .
\end{equation}%
By further restoring the spin wave velocity $c_{s}$, we arrive at the
\textquotedblleft Maxwell\textquotedblright \ term on the rhs of Eq. (\ref%
{effectiveL}) which describes the fluctuation ${a}_{\mu }^{h}$, with the
coefficients given by
\begin{equation}
\, \gamma =\frac{\vartheta }{12 c_{s}}\,,\text{ \  \  \ }\kappa =\frac{%
\vartheta c_{s}\pi}{12 }\,,  \label{g}
\end{equation}%
in which the dimensionless coefficient $\vartheta $ depends on the
dimensionless quantity $\frac{c_{s}\pi \, \delta \,}{m_{s}}$ as follows
\begin{equation}
\vartheta =\frac{\int_{0}^{\infty }\frac{dT}{T^{1/2}}e^{-T}\int_{0}^{1}d\ell
\,{(1-2\ell )\cdot {\pi c_{s}\, \delta \,T}m_{s}^{-2}\sinh \left( (1-2\ell ){%
\pi c_{s}\, \delta \,T}m_{s}^{-2}\right) }\, \cdot \, \left( \sinh ({\pi
c_{s}\, \delta \,T}m_{s}^{-2})\right) ^{-2}}{\int_{0}^{\infty }\frac{dT}{%
T^{1/2}}e^{-T}\int_{0}^{1}d\ell \,(1-2\ell )^{2}}  \label{theta}
\end{equation}%
Note that $\vartheta =1$ in the limit of the absence of \textquotedblleft
magnetic flux\textquotedblright $\,$in $\bar{A}^{h}$.

\section{Derivation details of topological order parameter $\mathcal{I}$}

\label{appendix_i}

To obtain $\mathcal{I}$, we first note that the Lagrangian with holon and
spinon currents circulating on loops $\mathcal{C}_{1, 2}$ ($m^*=(2t_h)^{-1}$%
)
\begin{equation}
L_{\text{SC}}\left[\mathcal{C}_1, \mathcal{C}_2\right]=\frac{\delta }{2{m^*}}%
\left(A_{\alpha }^s\right)^2+\frac{1}{2u_1}\left(A_0^s\right)^2+ \frac{i}{%
\pi }\epsilon _{\mu \nu \lambda }a_{\mu }^h\partial _{\nu }A_{\lambda }^s+%
\frac{1}{2}\frac{g_{\mu }}{\pi ^2}\left(\epsilon _{\mu \nu \lambda }\partial
_{\nu }a_{\lambda }^h\right)^2-i A_{\mu }^sJ_{\mu }^{\mathcal{C}_1}-i a_{\mu
}^hJ_{\mu }^{\mathcal{C}_2}
\end{equation}
is quadratic in the gauge fields thus allows rigorously integrating out $A^s$
and $a^h$ in the generating functional
\begin{equation}
\left \langle \hat{W}^s\left[\mathcal{C}_1\right]\hat{W}^h\left[\mathcal{C}_2%
\right]\right \rangle =\int D[ A^s a^h]e^{-\int d^3x L_{\text{SC}}\left[%
\mathcal{C}_1, \mathcal{C}_2\right]} \\
=\text{Const}.\times \int D[ a^h]\exp \left \{-\int d^3x\left(\frac{G_{\mu }%
}{2\pi ^2}\left(\epsilon _{\mu \nu \lambda }\partial _{\nu }a_{\lambda
}^h\right)^2-i a_{\mu }^hJ_{\mu }\right)\right \}
\end{equation}
in which we have defined
\begin{gather*}
G_{\mu }=g_{\mu }+\left(u_1,\frac{{m^*}}{\delta },\frac{{m^*}}{\delta }%
\right)_{\mu },\quad J_0=J_0^{\mathcal{C}_2}-\frac{i}{\pi }\frac{{m^*}}{%
\delta }\left(\partial _1J_2^{\mathcal{C}_1}-\partial _2J_1^{\mathcal{C}%
_1}\right), \\
J_1=J_1^{\mathcal{C}_2}-\frac{i}{\pi }\left(u_1\partial _2J_0^{\mathcal{C}%
_1}-\frac{{m^*}}{\delta }\partial _0J_2^{\mathcal{C}_1}\right),\quad
J_2=J_2^{\mathcal{C}_2}-\frac{i}{\pi }\left(\frac{{m^*}}{\delta }\partial
_0J_1^{\mathcal{C}_1}-u_1\partial _1J_0^{\mathcal{C}_1}\right)
\end{gather*}
and absorbed terms unrelated to mutual correlation between $\mathcal{C}_1$
and $\mathcal{C}_2$ into Const. which will be canceled by the denominator in
the definition of $\mathcal{I}$.

Adopting the ``deformed'' Lorentz gauge condition,
\begin{equation}
\frac{1}{G_0}\partial _0a_0^h+\frac{1}{G_1}\partial _1a_1^h+\frac{1}{G_2}%
\partial _2a_2^h=0
\end{equation}
and integrating out $a^h$ in the Feynman gauge\cite{Greiner96}, we get
\begin{equation}
\left \langle \hat{W}^s\left[\mathcal{C}_1\right]\hat{W}^h\left[\mathcal{C}_2%
\right]\right \rangle = \\
\text{Const}.\times \exp \left \{-\frac{1}{2}\frac{\pi ^2}{G_0G_1G_2}\int
d^3x\int d^3y\sum _{\mu } G_{\mu }J_{\mu }(x)J_{\mu }(y)f(x,y)\right \}
\label{eq:6}
\end{equation}
in which
\begin{equation*}
f(x,y)\equiv \int \frac{d^3k}{(2\pi )^3} \left(\sum _{\lambda } \frac{%
k_{\lambda }^2}{G_{\lambda }}\right)^{-1}e^{-i (x-y)\cdot k}
\end{equation*}

Expanding the integrand in Eq. \ref{eq:6} and extracting mutual correlation
terms between $\mathcal{C}_1$ and $\mathcal{C}_2$ only, and noting that $%
\int d^3x J_{\mu }^{\mathcal{C}_i}(x)=\oint _{\mathcal{C}_i}dx_{\mu }$, $%
i=1, 2$, we get
\begin{equation}
\begin{split}
i\mathcal{I}=&-\frac{1}{2}\frac{\pi ^2}{G_0G_1G_2}\frac{2i}{\pi } \bigg(%
u_1\oint _{\mathcal{C}_1}dy_0\left(G_1\oint _{\mathcal{C}_2}dx_1\partial
_2^yf(x,y)-G_2\oint _{\mathcal{C}_2}dx_2\partial _1^yf(x,y)\right) \\
&+\frac{{m^*}}{\delta }\oint _{\mathcal{C}_1}dy_1\left(G_2\oint _{\mathcal{C}%
_2}dx_2\partial _0^yf(x,y)-G_0\oint _{\mathcal{C}_2}dx_0\partial
_2^yf(x,y)\right) \\
&+\frac{{m^*}}{\delta }\oint _{\mathcal{C}_1}dy_2\left(G_0\oint _{\mathcal{C}%
_2}dx_0\partial _1^yf(x,y)-G_1\oint _{\mathcal{C}_2}dx_1\partial
_0^yf(x,y)\right)\bigg)
\end{split}%
\end{equation}

Applying Stokes' formula to transform the loop integral over $\mathcal{C}_1$
into a surface integral over a disk $D_1$ such that $\partial D_1=\mathcal{C}%
_1$, and noting that $\oint _{\mathcal{C}_2}dx_{\mu }\partial _{\mu
}^xh(x)=0 $, we get
\begin{equation}
\begin{split}
i\mathcal{I}=&-\frac{1}{2}\frac{\pi ^2}{G_0G_1G_2}\frac{2i}{\pi }\bigg(\int
_{D_1}dy_1\wedge dy_2\oint _{\mathcal{C}_2}dx_0\frac{{m^*}}{\delta }G_0\int
\frac{d^3k}{(2\pi )^3}\left(-k_1^2-k_2^2\right) \left(\sum _{\lambda } \frac{%
k_{\lambda }^2}{G_{\lambda }}\right)^{-1}e^{-i (x-y)\cdot k} \\
&+\int _{D_1}dy_0\wedge dy_1\oint _{\mathcal{C}_2}dx_2G_2\int \frac{d^3k}{%
(2\pi )^3}\left(-u_1k_1^2-\frac{{m^*}}{\delta }k_0^2\right) \left(\sum
_{\lambda } \frac{k_{\lambda }^2}{G_{\lambda }}\right)^{-1}e^{-i (x-y)\cdot
k} \\
&+\int _{D_1}dy_2\wedge dy_0\oint _{\mathcal{C}_2}dx_1G_1 \int \frac{d^3k}{%
(2\pi )^3}\left(-\frac{{m^*}}{\delta }k_0^2-u_1k_2^2\right) \left(\sum
_{\lambda } \frac{k_{\lambda }^2}{G_{\lambda }}\right)^{-1}e^{-i (x-y)\cdot
k}\bigg)
\end{split}
\label{eq:10}
\end{equation}

Given that $G_1=G_2$, Eq. \ref{eq:10} is greatly reduced if $\mathcal{C}_1$
lies within a spatial plane,
\begin{equation}
\begin{split}
\mathcal{I}&=-\frac{1}{2}\frac{\pi ^2}{G_0G_1G_2}\frac{2}{\pi }\int
_{D_1}dy_1\wedge dy_2\oint _{\mathcal{C}_2}dx_0\frac{{m^*}}{\delta }G_0 \int
\frac{d^3k}{(2\pi )^3}\left(-k_1^2-k_2^2-\frac{G_1}{G_0}k_0^2\right)
\left(\sum _{\lambda } \frac{k_{\lambda }^2}{G_{\lambda }}\right)^{-1}e^{-i
(x-y)\cdot k} \\
&=\frac{1}{2}\frac{\pi ^2}{G_0G_1G_2}\frac{2}{\pi }\frac{{m^*}}{\delta }%
G_0G_1\int _{D_1}dy_1\wedge dy_2\oint _{\mathcal{C}_2}dx_0\delta ^{(3)}(x-y)
\\
&=\left(\lambda_0/{\lambda_p}\right)^2\pi \Theta
\end{split}%
\end{equation}
in which $\Theta$ is the Gaussian linking number of $\mathcal{C}_1$ and $%
\mathcal{C}_2$ loops. To obtain the last line of above identity, the
definition of $\lambda_p$ and $\lambda_0$ in Eq. (\ref{lambda-p}) is applied.

\end{document}